\documentclass[preprint,tightenlines,aps,prd,groupedaddress,showpacs]
              {revtex4}%
\usepackage{graphicx}
\usepackage{amsmath}
\usepackage{bm}

\begin{document}

\preprint{\begin{tabular}{l}
          FERMILAB-PUB-02/303-T\\
          ANL-HEP-PR-02-105\end{tabular}}
\title{\mbox{}\\[10pt]
Exclusive Double-Charmonium Production \\
from $\bm{e^+ e^-}$ Annihilation into Two Virtual Photons}


\author{Geoffrey T.~Bodwin}
\affiliation{
High Energy Physics Division, 
Argonne National Laboratory, 
9700 South Cass Avenue, Argonne, Illinois 60439}

\author{Eric Braaten}
\affiliation{
Physics Department, Ohio State University, Columbus, Ohio 43210}
\affiliation{
Fermi National Accelerator Laboratory, P.~O.~Box 500, Batavia, Illinois 60510}

\author{Jungil Lee}
\affiliation{
Department of Physics, Korea University,
Seoul 136-701, Korea}


\date{\today}
\begin{abstract}
We calculate the contributions from QED processes involving two virtual
photons to the cross sections for $e^+ e^-$ annihilation into two
charmonium states with the same C-parity. Generically, the cross
sections are three orders of magnitude smaller than those for charmonia
with opposite C-parity because they are suppressed by a factor of
$\alpha^2/\alpha_s^2$. However, if both charmonia have quantum numbers
$J^{PC} = 1^{--}$, then there is a contribution to the cross section
that involves the fragmentation of each photon into a charmonium. The
fragmentation contribution is enhanced by powers of $E_{\rm
beam}/m_c$, the ratio of the beam energy to the charm-quark mass, and
this enhancement can compensate for the suppression factor that is
associated with the coupling constants. In particular, the predicted
cross section for $J/\psi + J/\psi$ at the $B$ factories is larger than
that for $J/\psi + \eta_c$.
\end{abstract}



\maketitle


\section{Introduction
	\label{intro}}

Recently, the Belle Collaboration observed $e^+ e^-$ annihilation into
two charmonium states at a center-of-mass energy $\sqrt{s} = 10.6$ GeV
by studying the recoil momentum spectrum of the $J/\psi$
\cite{Abe:2002rb}. The collaboration measured the cross section for
$J/\psi + \eta_c$ and also found evidence for production of $J/\psi +
\chi_{c0}$ and $J/\psi + \eta_c(2S)$. With more precise measurements and
additional data, it may be possible to discover the $D$-wave charmonium
states $\eta_{c2}(1D)$ and $\psi_2(1D)$ at the $B$ factories by studying
the momentum spectra of the $J/\psi$ and other charmonium states.

The exclusive cross sections for double-charmonium states provide
stringent tests of the NRQCD factorization method for calculating the
cross sections for heavy-quarkonium production \cite{BBL}. The
predictions of the NRQCD factorization method for exclusive 
double-charmonium production 
reduce essentially to those of the color-singlet model for
quarkonium production~\cite{CSM}, up to relativistic corrections that
are second order in the relative velocity $v$ of the quark and
antiquark in the 
charmonium. The nonperturbative factors in the NRQCD
factorization formula for the cross section in the nonrelativistic limit
can be determined phenomenologically from electromagnetic annihilation
decay rates. Cross sections for double-charmonium production can
therefore be predicted, up to corrections that are suppressed by powers
of $v^2$, without any unknown phenomenological factors.

Recent calculations of the cross section for $J/\psi + \eta_c$ have
given results that are about an order of magnitude smaller than the
Belle measurement \cite{Braaten:2002fi,Liu:2002wq}. The relativistic
corrections are large, and they may account for part of the discrepancy
\cite{Braaten:2002fi}. It has also been suggested that there may be large 
nonperturbative corrections to double-charmonium cross sections at 
the $B$-factory energy \cite{Liu:2002wq}.

In this paper, we calculate the cross sections for $e^+ e^-$ annihilation
into two charmonium states that have the same charge-conjugation parity
(C-parity), such as $J/\psi+J/\psi$. These processes proceed, at leading
order in the QCD coupling $\alpha_s$, through QED diagrams with that
contain two virtual photons. One might expect that these cross
sections would be much smaller than those for charmonia with opposite
C-parity because they are suppressed by a factor of
$\alpha^2/\alpha_s^2$. However, if both charmonia have quantum numbers
$J^{PC} = 1^{--}$, then there is a contribution to the cross section in
which each photon fragments into a charmonium~\cite{Bodwin:2002fk}.
The fragmentation contribution is enhanced by powers of $E_{\rm
beam}/m_c$, where $E_{\rm beam}$ is the beam energy and $m_c$ is the
charm-quark mass \cite{Bodwin:2002fk}. This enhancement can
compensate for the suppression factor that is associated with the
coupling constants. In particular, the predicted cross section for
$J/\psi+J/\psi$ at the $B$ factories is larger than that for $J/\psi +
\eta_c$.

\section{ Cross sections
	\label{sec:CSM}}

In this section, we calculate the contributions from $e^+ e^-$
annihilation through two virtual photons to the production cross
sections for double-charmonium states $H_1+H_2$. Charge-conjugation
symmetry requires the two charmonia to be either two $C=-1$ states or two
$C=+1$ states. We express our results in terms of the ratio $R[H_1+H_2]$,
which is defined by
\begin{eqnarray}
R[H_1 + H_2] = 
\frac{\sigma[e^+ e^- \to H_1 + H_2]}{ \sigma[e^+ e^- \to \mu^+ \mu^-]},
\label{R-def}
\end{eqnarray}
where $\sigma[e^+ e^- \to \mu^+ \mu^-]=\pi\alpha^2/(3E_{\rm beam}^2)$ and
$E_{\rm beam}=\sqrt{s}/2$ is the beam energy in the center-of-momentum (CM)
frame. We give the angular distributions $dR/dx$ for each of the
helicity states of $H_1$ and $H_2$. The angular variable is $x = \cos
\theta$, where $\theta$ is the angle between the $e^-$ and $H_1$ in the
$e^+e^-$ CM frame.

\subsection{Asymptotic behavior}

\begin{figure}
\includegraphics[width=12cm,angle=-90]{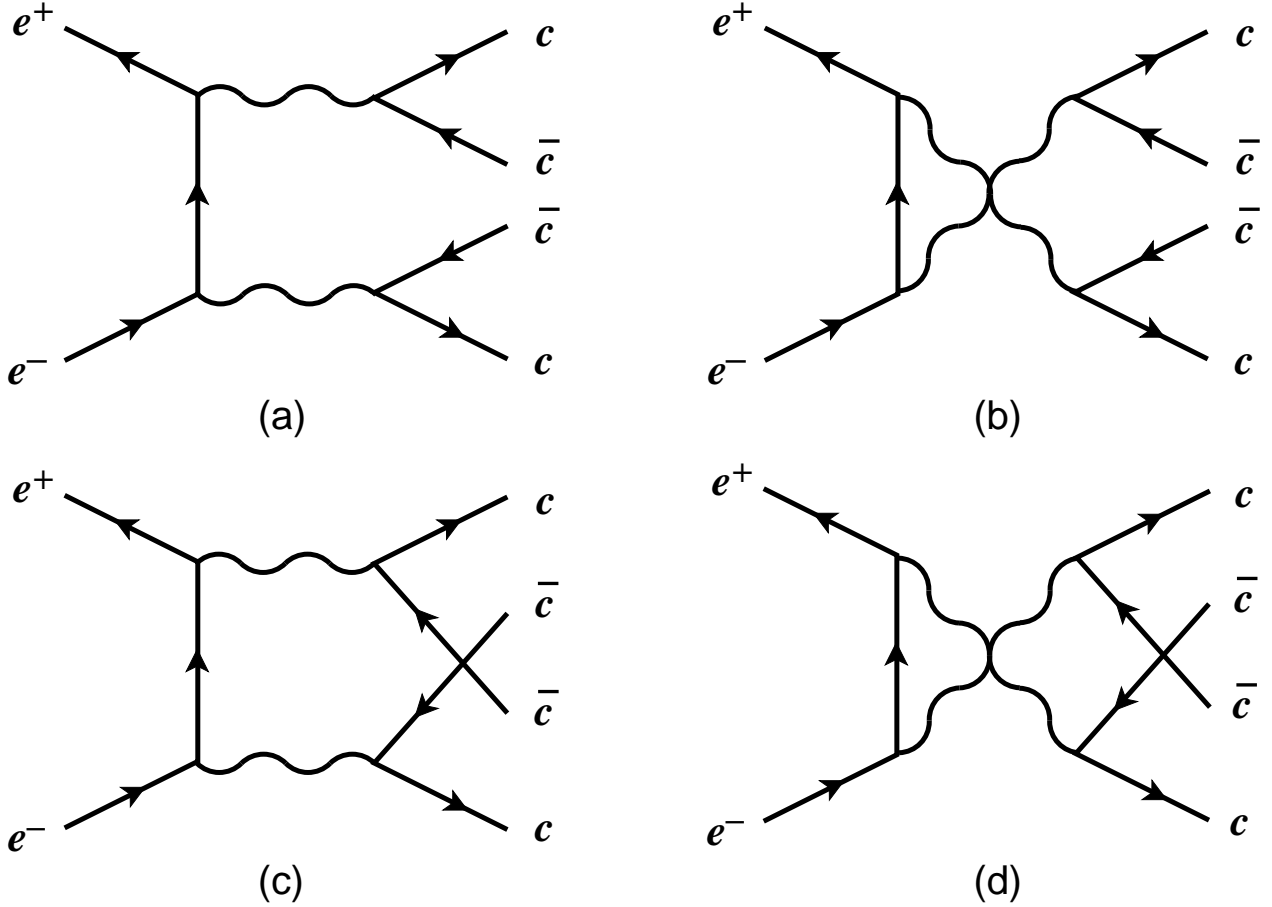}
\caption{\label{fig1}%
QED diagrams for the process $e^+ e^- \to \gamma^* \gamma^* \to c \bar
c_1 + c \bar c_1$. The upper and lower $c\bar{c}$ pairs evolve into
$H_1$ and $H_2$, respectively.}
\end{figure}

The 4 QED diagrams that contribute to the process $e^+ e^- \to \gamma^*
\gamma^* \to c \bar c_1 + c \bar c_1$ at order $\alpha^4$ are shown in
Fig.~\ref{fig1}. The diagrams in Figs.~\ref{fig1}(a) and \ref{fig1}(b)
contribute only for charmonia with quantum numbers $J^{PC} = 1^{--}$,
such as $J/\psi$, $\psi(2S)$, and $\psi_1(1D)$. The procedure for
calculating the matrix element for $e^+ e^- \to H_1(P_1) + H_2(P_2)$
from the matrix element for $e^+ e^- \to c(p_1) \bar c(\bar p_1) +
c(p_2) \bar c(\bar p_2)$ is summarized in Ref.~\cite{Braaten:2002fi}.

When the 
energy $E_{\rm beam}$ is much larger than the charm-quark mass
$m_c$, the relative sizes of the double-charmonium cross sections are
governed largely by the number of kinematic suppression factors $r^2$,
where the variable $r$ is defined by
\begin{eqnarray}
r^2 &=& \frac{4 m_c^2 }{ E_{\rm beam}^2}.
\label{r-def}
\end{eqnarray}
If we set $m_c = 1.4$ GeV and $E_{\rm beam} = 5.3$ GeV, then $r^2 =
0.28$.

The asymptotic behavior of the ratio $R[H_1 + H_2]$ as $r \to 0$ can be
determined from the helicity selection rules for exclusive processes in
perturbative QCD \cite{Chernyak:dj,Brodsky:1981kj}, which apply equally
well to QED. For generic charmonia, whose quantum numbers are
different from those of the photon, only the diagrams in
Figs.~\ref{fig1}(c) and \ref{fig1}(d) contribute. There is a suppression
factor $r^2$ for each $c \bar c$ pair with small relative momentum in
the final state. Thus the ratio $R[H_1(\lambda_1) + H_2(\lambda_2)]$
decreases at least as fast as $r^4$ as $r \to 0$. This slowest
asymptotic decrease can occur only if the sum of the helicities of the
charmonia vanishes: $\lambda_1 + \lambda_2 = 0$. For every unit of
helicity by which this selection rule is violated, there is a further
suppression factor of $r^2$. Thus the estimate for the ratio $R$ at
leading-order in the QED and QCD coupling constants is
\begin{eqnarray}
R[H_1(\lambda_1) + H_2(\lambda_2)]
\; \sim \; \alpha^2 (v^2)^{3+L_1+L_2} (r^2)^{2+|\lambda_1+\lambda_2|}.
\label{R:lam1+lam2}
\end{eqnarray}
The factor of $v^{3+2L}$ for a charmonium state with orbital angular
momentum $L$ comes from the NRQCD matrix elements. At leading order,
there may, of course, be further suppression factors of $r^2$, and the
cross section could even vanish. For charmonia with opposite charge
conjugation, the estimate for $R$ is given by Eq.~(\ref{R:lam1+lam2})
with $\alpha^2$ replaced by $\alpha_s^2$. Thus, the cross sections for
generic charmonia with the same charge conjugation are suppressed
compared to those for opposite charge conjugation by a factor of
$\alpha^2/\alpha_s^2 \approx 10^{-3}$.

Production cross sections for charmonia with $J^{PC} = 1^{--}$ are an
exception to this generic estimate because they also receive
contributions from the diagrams in Figs.~\ref{fig1}(a) and \ref{fig1}(b),
which we will refer to as photon-fragmentation diagrams. These
contributions are enhanced because the virtual-photon propagators are of
order $1/m_c^2$ instead of order $1/E_{\rm beam}^2$. In the amplitude,
there are also two numerator factors of $m_c$ instead of $E_{\rm beam}$,
which arise from the $c\bar c$ electromagnetic currents. Hence, the net
enhancement of the squared amplitude is $(E_{\rm beam}/m_c)^4$, and the
contributions to $R$ can be nonzero in the limit $r\to 0$. As $r \to
0$ with fixed scattering angle $\theta$, the photon-fragmentation
contributions to the cross section factor into the cross section for
$e^+ e^- \to \gamma \gamma$ with photon scattering angle $\theta$ and
the fragmentation probabilities for $\gamma \to H_1$ and $\gamma \to
H_2$. These fragmentation probabilities are nonzero at order $\alpha$
only for $J^{PC} = 1^{--}$ states with helicities satisfying $\lambda_1
= -\lambda_2 = \pm 1$. The contribution to the ratio $R$ for $J/\psi +
J/\psi$ has the behavior
\begin{eqnarray}
R[J/\psi(\lambda_1) + J/\psi(\lambda_2)]
\; \sim \; \alpha^2 (v^2)^{3},  
\hspace{1cm} \lambda_1 = -\lambda_2 = \pm 1.
\label{Rest:psi+psi}
\end{eqnarray}
This estimate applies equally well to $J/\psi + \psi(2S)$ and $\psi(2S)
+ \psi(2S)$. For $J/\psi + \psi_1(1D)$ and $\psi(2S) + \psi_1(1D)$,
there is an additional suppression factor of $v^4$ from the NRQCD matrix
elements.  For $\psi_1(1D) + \psi_1(1D)$, the additional suppression
factor is $v^8$.

Using the QCD version of Eq.~(\ref{R:lam1+lam2}), we obtain the 
estimate 
\begin{eqnarray}
R[J/\psi + \eta_c]
\; \sim \; \alpha_s^2 (v^2)^{3} (r^2)^3,
\label{Rest:psi+eta}
\end{eqnarray}
which holds generally for $S$-wave final states with opposite C-parity.
The ratio $R$ in Eq.~(\ref{Rest:psi+psi}) is suppressed relative to
Eq.~(\ref{Rest:psi+eta}) by a factor of $(\alpha/\alpha_s)^2 \approx
10^{-3}$, but the enhancement factor that scales as $r^{-6}$ makes the
cross sections comparable in magnitude at the energy of a $B$ factory.

\subsection{$\bm{S}$-wave + $\bm{S}$-wave}

The cross section for $\eta_c + \eta_c$ receives contributions only from
the nonfragmentation diagrams in Figs.~\ref{fig1}(c) and \ref{fig1}(d).
The angular distribution is
\begin{eqnarray}
\frac{dR}{dx}[\eta_c + \eta_c] =
\frac{ \pi^2 e_c^4 \alpha^2 }{3} \,
x^2(1-x^2)\,
r^4 (1-r^2)^{5/2}\,
\frac{\langle O_1 \rangle_{\eta_c} \langle O_1 \rangle_{\eta_c} }{ m_c^6}.
\label{dRdx:eta+eta}
\end{eqnarray}
The total ratio $R$ for $\eta_c + \eta_c$ is obtained by integrating
over $x$ only from 0 to 1 in order to avoid double-counting of the
identical final-state particles:
\begin{eqnarray}
R[\eta_c + \eta_c] = 
\frac{ 2\pi^2 e_c^4 \alpha^2 }{ 45} \,
r^4 (1-r^2)^{5/2} \,
\frac{\langle O_1 \rangle_{\eta_c} \langle O_1 \rangle_{\eta_c} }{ m_c^6}.
\label{R:eta+eta}
\end{eqnarray}
Note that the ratio (\ref{R:eta+eta}) 
depends on the charm-quark mass $m_c$ explicitly and also
through the variable $r$ defined in Eq.~(\ref{r-def}).
One of the factors of $(1-r^2)^{1/2}$ in Eq.~(\ref{R:eta+eta})
is the nonrelativistic limit of the phase-space factor
 $P_{\rm CM}/E_{\rm beam}$,
where $P_{\rm CM}$ is the momentum of either charmonium 
in the center-of-momentum frame.  It can be expressed as
\begin{eqnarray}
\frac{ P_{\rm CM} }{ E_{\rm beam}}
= \frac{\lambda^{1/2}(s,M_{H_1}^2,M_{H_2}^2) }{ s},
\label{P_cm}
\end{eqnarray}
where $\lambda(x,y,z) = x^2 + y^2 + z^2 -2(xy + yz + zx)$ and
$M_{H_i}$ is the mass of $H_i$.

The angular distribution $dR/dx$ for $\eta_c + \eta_c(2S)$ is given by
an expression identical to Eq.~(\ref{dRdx:eta+eta}), except that one of
the factors of $\langle O_1 \rangle_{\eta_c}$ is replaced by $\langle
O_1 \rangle_{\eta_c(2S)}$. Since the two final-state particles are
distinguishable, the total ratio $R$ for $\eta_c + \eta_c(2S)$ is
obtained by integrating over $x$ from $-1$ to $1$. Thus, the total
ratio $R$ for $\eta_c + \eta_c(2S)$ is obtained from
Eq.~(\ref{R:eta+eta}) by replacing one of the factors of $\langle O_1
\rangle_{\eta_c}$ by $\langle O_1 \rangle_{\eta_c(2S)}$ and multiplying
by a factor 2.

The cross section for $J/\psi + J/\psi$ receives contributions from
all four diagrams in Fig.~\ref{fig1}, 
including the photon-fragmentation diagrams.
The angular distribution $dR/dx$ is
\begin{eqnarray}
\frac{dR}{dx}[J/\psi(\lambda_1) + J/\psi(\lambda_2)] = 
\frac{\pi^2 e_c^4 \alpha^2 }{3} \,
\frac{F(\lambda_1,\lambda_2,x)(1-r^2)^{1/2}}
     {[4(1-r^2)(1-x^2)+r^4]^2} \,
\frac{\langle O_1 \rangle_{J/\psi} \langle O_1 \rangle_{J/\psi} }{ m_c^6},
\label{dRdx:psi+psi}
\end{eqnarray}
where the entries of $F(\lambda_1,\lambda_2,x)$ are
\begin{subequations}
\begin{eqnarray}
F(0,0,x)&=&r^4 x^2 (1-x^2) [ 2 + 3 r^4 -r^6 +4 x^2 (1-r^4) ]^2,
\\
F(\pm 1,\mp 1,x)&=& (1+x^2) (1-x^2) 
[ 6 - 7 r^2 +4 r^4-r^6 +4 x^2 r^2 (1-r^2) ]^2,
\label{Fpm}
\\
F(\pm 1,\pm 1,x)&=& r^4 x^2 (1-x^2) 
[3 - 4 r^2 +4 r^4-r^6 +4 x^2 r^2 (1-r^2)  ]^2,
\\
F(\pm 1,0,x)&=&F(0,\pm 1,x)= (r^2/2) [ 
  (2-r^2)^2 (3- 2 r^2 + r^4)^2
-x^2 ( 72 - 228 r^2 
\nonumber\\ &&
+ 359 r^4 - 292 r^6 + 130 r^8 -  32 r^{10} + 3 r^{12} )
+4 x^4 ( 9 - 60 r^2 +122 r^4 
\nonumber\\ &&
-112 r^6 + 58 r^8 - 14 r^{10} +    r^{12} )
+16 x^6 r^2 (1-r^2) ( 6- 11 r^2+ 11 r^4
\nonumber\\ &&
- 2 r^6)
+64 x^8 r^4 (1-r^2)^2
].
\label{F10}
\end{eqnarray}
\end{subequations}
The total ratio $R$ is obtained by summing over the helicities
$\lambda_1$ and $\lambda_2$ and integrating over $x$ only from $0$ to
$1$. The resulting expression is so complicated that it is not very
useful to write it down. The ratio  $dR/dx$ for $J/\psi + \psi(2S)$ is
given by an expression identical to Eq.~(\ref{dRdx:psi+psi}), except
that one of the factors of $\langle O_1 \rangle_{J/\psi}$ is replaced by
$\langle O_1 \rangle_{\psi(2S)}$ and the range of $x$ is from $-1$ to
$1$.

In the limit $r \to 0$ with $x$ fixed, only the transverse helicity states
$\lambda_1 = - \lambda_2 = \pm 1$ in Eq.~(\ref{Fpm}) contribute. 
The angular distribution $dR/dx$ summed over helicities
reduces in this limit to
\begin{eqnarray}
\frac{dR}{dx}[J/\psi(\pm 1) + J/\psi(\mp 1)] \approx
\frac{3(1+x^2)}{2(1-x^2)} \, P_{\gamma \to J/\psi}^2,
\label{dRdx:frag}
\end{eqnarray}
where $P_{\gamma \to J/\psi}$ is the probability for a photon to fragment 
into a $J/\psi$ \cite{Fleming:1994iu}:
\begin{eqnarray}
P_{\gamma \to J/\psi} 
= e_c^2 \pi \alpha \frac{\langle O_1 \rangle_{J/\psi}}{ m_c^3}.
\label{P-gampsi}
\end{eqnarray}
The coefficient of $P_{\gamma \to J/\psi}^2$ in Eq.~(\ref{dRdx:frag})
is $dR/dx$ for the process $e^+ e^- \to \gamma \gamma$. The differential
ratio (\ref{dRdx:frag}) is compatible with the asymptotic form
(\ref{Rest:psi+psi}). However the integrated cross section is enhanced
by a factor $\ln(8/r^4)$, relative to Eq.~(\ref{Rest:psi+psi}). The
logarithmic factor arises because, as can be seen from
Eq.~(\ref{dRdx:psi+psi}), the potential divergence at $x=\pm 1$ in the
integral of Eq.~(\ref{dRdx:frag}) is cut off at $1\pm x \sim r^4/8$.

Chang, Qiao, and Wang have pointed out that the cross sections  for $e^+
e^- \to J/\psi + \gamma$ is also large if $E_{\rm beam} \gg m_c$
\cite{Chang:1997dw}. The reason is that this process has a fragmentation
contribution that factors into the short-distance cross section for $e^+
e^- \to \gamma \gamma$ and a single fragmentation probability $P_{\gamma
\to J/\psi}$.  The ratio $dR/dx$ for this process therefore has a finite
limit as $r \to 0$, and the cross section is sharply peaked near the
beam direction.

\subsection{$\bm{S}$-wave + $\bm{P}$-wave}
The cross section for $J/\psi + h_c$ receives contributions only from
the nonfragmentation diagrams in Figs.~\ref{fig1}(c) and \ref{fig1}(d).
The angular distribution is
\begin{eqnarray}
\frac{dR}{dx}[J/\psi(\lambda_1) + h_c(\lambda_2)] =
\frac{ \pi^2 e_c^4 \alpha^2 }{24} \,
G(\lambda_1,\lambda_2,x)\, r^4 (1-r^2)^{3/2} \,
\frac{\langle O_1 \rangle_{J/\psi} \langle O_1 \rangle_{h_c} }{ m_c^8} ,
\label{dRdx:psi+hc}
\end{eqnarray}
where the non-zero entries of $G(\lambda_1,\lambda_2,x)$ are
\begin{subequations}
\begin{eqnarray}
G(0,0,x)         &=& 2 (1-x^2),
\\
G(0,\pm 1,x)     &=& r^2 (1- 3x^2 + 4 x^4),
\\
G(\pm 1,\mp 1,x) &=& 8 (1-x^4),
\\
G(\pm 1,0,x)     &=& 4 r^2 x^2 (1+x^2).
\end{eqnarray}
\label{G-def}
\end{subequations}
The total ratio $R$ is obtained by summing over the helicities 
$\lambda_1$ and $\lambda_2$ and integrating over $x$ from $-1$ to $1$:
\begin{eqnarray}
R[J/\psi + h_c] =
\frac{ \pi^2 e_c^4 \alpha^2 }{45} \,
r^4  (53+22 r^2) (1-r^2)^{3/2} \,
\frac{\langle O_1 \rangle_{J/\psi} \langle O_1 \rangle_{h_c} }{ m_c^8} .
\label{R:psi+hc}
\end{eqnarray}

The cross section for $\eta_c+\chi_{cJ}$ receives contributions only from
the nonfragmentation diagrams in Figs.~\ref{fig1}(c) and \ref{fig1}(d).
The angular distribution is
\begin{eqnarray}
\frac{dR}{dx}[\eta_c + \chi_{cJ}(\lambda_2)] =
\frac{2 \pi^2 e_c^4 \alpha^2 }{27} \,
G_J(\lambda_2,x)\, r^4 (1-r^2)^{3/2} \,
\frac{\langle O_1 \rangle_{\eta_c} \langle O_1 \rangle_{\chi_{cJ}} }{ m_c^8} ,
\label{dRdx:eta+chi}
\end{eqnarray}
where the non-zero entries of $G_J(\lambda_2,x)$ are
\begin{subequations}
\begin{eqnarray}
G_0(0,x)         &=& (3/4) (1-x^2),
\\
G_1(0,x)     &=& 9 x^2 (1- x^2),
\\
G_1(\pm 1,x)     &=&(9r^2/32)(1+x^2 + 16 x^4),
\\
G_2(0,x) &=& (3/4)(1-x^2),
\\
G_2(\pm 1,x)     &=& (9r^2/32)(1+ x^2).
\end{eqnarray}
\label{GJ-def}
\end{subequations}
The total ratio $R$ is obtained by summing over the helicities
$\lambda_1$ and $\lambda_2$ and integrating over $x$ from $-1$ to $1$:
\begin{eqnarray}
R[\eta_c + \chi_{cJ}] =
\frac{2 \pi^2 e_c^4 \alpha^2 }{27} \,
G_J\, r^4 (1-r^2)^{3/2} \,
\frac{\langle O_1 \rangle_{\eta_c} \langle O_1 \rangle_{\chi_{cJ}} }{ m_c^8} ,
\label{R:eta+chi}
\end{eqnarray}
where the coefficients $G_J$ are
\begin{subequations}
\begin{eqnarray}
G_0         &=& 1 ,
\\
G_1     &=&(3/10)(8+17r^2), 
\\
G_2     &=&(1/2)(2+3r^2).
\end{eqnarray}
\label{RGJ-def}
\end{subequations}

\subsection{$\bm{P}$-wave + $\bm{P}$-wave}
The cross section for $h_c + h_c$ receives contributions only from
the nonfragmentation diagrams in Figs.~\ref{fig1}(c) and \ref{fig1}(d).
The angular distribution is
\begin{eqnarray}
\frac{dR}{dx}[h_c(\lambda_1) + h_c(\lambda_2)] =
\frac{ \pi^2 e_c^4 \alpha^2 }{24} \,
H(\lambda_1,\lambda_2,x)\, r^4 (1-r^2)^{1/2} \,
\frac{\langle O_1 \rangle_{h_c} \langle O_1 \rangle_{h_c} }{ m_c^{10}} ,
\label{dRdx:hc+hc}
\end{eqnarray}
where the functions $H(\lambda_1,\lambda_2,x)$ are
\begin{subequations}
\begin{eqnarray}
H(0,0,x)         &=& 2 x^2 (1-x^2)[ 4 - 6r^2 + 3r^4 - 4x^2(1-r^2) ]^2,
\\
H(\pm1,\mp 1,x)  &=& h_1(x) + h_1(-x),
\label{H11}
\\
H(\pm1 ,0,x)     &=& H(0,\pm1,x) = h_2(x) + h_2(-x),
\label{H10}
\\
H(\pm1,\pm 1,x)  &=& (r^4/2) x^2(1-x^2) [ 1+(5-4x^2)(1-r^2) ]^2.
\end{eqnarray}
\label{H-def}
\end{subequations}
The functions $h_i(x)$ in Eqs.~(\ref{H11}) and (\ref{H10}) are defined by
\begin{subequations}
\begin{eqnarray}
h_1(x)&=&
(1/4)(1+x)^2(1-x^2) [  (2-r^2)^2+4x(1-x)r^2(1-r^2)  ]^2,
\\
h_2(x)&=&
(r^2/8)(1-x)^2[  (1+2x)(2-r^2) - 8x^2(1+x)(1-r^2)  ]^2.
\end{eqnarray}
\label{hi-def}
\end{subequations}
The total ratio $R$ is obtained by summing over the helicities 
$\lambda_1$ and $\lambda_2$ and integrating over $x$ from $0$ to $1$:
\begin{eqnarray}
R[h_c + h_c] &=&
\frac{ \pi^2 e_c^4 \alpha^2 }{3780} \,
(2272 - 3880 r^2+ 3300 r^4 - 2158 r^6 + 781 r^8)
\nonumber\\
&&\times\;r^4 (1-r^2)^{1/2} \,
\frac{\langle O_1 \rangle_{h_c} \langle O_1 \rangle_{h_c} }{ m_c^{10}} .
\label{R:hc+hc}
\end{eqnarray}

We have not calculated the cross sections for $\chi_{cJ} + \chi_{cJ'}$. 
They receive contributions only from the nonfragmentation diagrams in 
Figs.~\ref{fig1}(c) and \ref{fig1}(d).
We therefore expect them to be comparable to 
the cross section for $h_c + h_c$.   

\subsection{$\bm{S}$-wave + $\bm{D}$-wave}
The cross section for $\eta_c + \eta_{c2}$ receives contributions only from
the nonfragmentation diagrams in Figs.~\ref{fig1}(c) and \ref{fig1}(d).
The angular distribution is
\begin{eqnarray}
\frac{dR}{dx}[\eta_c + \eta_{c2}(\lambda_2)] =
\frac{ \pi^2 e_c^4 \alpha^2 }{288} \,
I(\lambda_2,x)\, r^4 (1-r^2)^{1/2} \,
\frac{\langle O_1 \rangle_{\eta_c} \langle O_1 \rangle_{\eta_{c2}} }
     { m_c^{10}} ,
\label{dRdx:eta+eta_c2}
\end{eqnarray}
where the functions $I(\lambda_2,x)$ are
\begin{subequations}
\begin{eqnarray}
I(0,x)         &=& 
x^2 (1-x^2) [  4-8 r^2+5 r^4 +4 x^2 (1-r^2) (2-r^2)   ]^2,
\\
I(\pm 1,x)  &=& i_1(x) + i_1(-x),
\label{I1}
\\
I(\pm 2,x)     &=&  i_2(x) + i_2(-x).
\label{I2}
\end{eqnarray}
\label{I-def}
\end{subequations}
The functions $i_i(x)$ in Eqs.~(\ref{I1}) and (\ref{I2}) are defined by
\begin{subequations}
\begin{eqnarray}
i_1(x)&=&
 (3r^2/4)  (1-x)^2 [ 2-3 r^2-2 x r^2 - 8 x^2 (1+x) (1-r^2) ]^2,
\\
i_2(x)&=&
 (3r^4/4)  (1-x^2) (1-x)^2 [1+(1-r^2) (1+2 x)^2]^2.
\end{eqnarray}
\label{ii-def}
\end{subequations}
Then total ratio $R$ is obtained by summing over the helicities 
$\lambda_2$ and integrating over $x$ from $-1$ to $1$:
\begin{eqnarray}
R[\eta_c + \eta_{c2}] &=&
\frac{ \pi^2 e_c^4 \alpha^2 }{22680} \,
(1232 - 3048 r^2 + 4880 r^4 - 3726 r^6 +1439 r^8 )
\nonumber\\
&&\times
\;r^4 (1-r^2)^{1/2} \,
\frac{\langle O_1 \rangle_{\eta_c} \langle O_1 \rangle_{\eta_{c2}} }
     { m_c^{10}}
.
\label{R:eta+eta_c2}
\end{eqnarray}

The cross section for $J/\psi+\psi_1$ receives contributions from all four
diagrams in Fig.~\ref{fig1}, 
including the photon fragmentation diagrams.
The angular distribution is
\begin{eqnarray}
\frac{dR}{dx}[J/\psi(\lambda_1) + \psi_1(\lambda_2)] =
\frac{ \pi^2 e_c^4 \alpha^2 }{5760} \,
\frac{J(\lambda_1,\lambda_2,x)\, (1-r^2)^{1/2} }
     {[4(1-r^2)(1-x^2)+r^4]^2}
\,
\frac{\langle O_1 \rangle_{J/\psi} \langle O_1 \rangle_{\psi_1} }{ m_c^{10}} ,
\label{dRdx:psi+psi_1}
\end{eqnarray}
where the functions $J(\lambda_1,\lambda_2,x)$ are
\begin{subequations}
\begin{eqnarray}
J(0,0,x)         &=&
2\, r^4\, x^2\, (1-x^2)\, \big[ 
84 + 124 r^2 - 158 r^4 + 101 r^6 - 22 r^8 
+ 4 x^2 (1-r^2) 
\nonumber\\&&\times
(34-37 r^2+34 r^4-2 r^6)
-32 x^4 (1-r^2)^2 (2-r^2)
\big]^2,
\\
J(\pm1,\mp 1,x)  &=& j_{+-}(x) + j_{+-}(-x),
\\
J(\pm1 ,0,x)     &=& j_{+0}(x) + j_{+0}(-x),
\\
J(0 ,\pm 1,x)    &=&  j_{0+}(x) + j_{0+}(-x),
\\
J(\pm 1,\pm 1,x) &=&  j_{++}(x) + j_{++}(-x).
\end{eqnarray}
\label{J-def}
\end{subequations}
The functions $j_{\lambda\lambda'}(x)$ are defined by
\begin{subequations}
\begin{eqnarray}
 j_{+-}(x) &=&
(1/4) (1-x^2) (1-x)^2 \big[
- 3 (2-r^2) ( 52 + 16 r^2 - 38 r^4 + 11 r^6 +  2 r^8 )
\nonumber\\&&
- 24 x   r^4 (2-r^2)^2 (1-r^2)
+ 4  x^2 r^2 (1-r^2) (40 -85 r^2 +22 r^4 - 6 r^6)
\nonumber\\&&
+ 96 x^3 r^4 (1-r^2)^2
- 32 x^4 r^2 (1-r^2)^2 (2-3 r^2)
\big]^2
,
\\
 j_{+0}(x) &=&
(r^2/8) (1-x)^2 \big[
   64 x^5 r^2 (1-r^2)^2
+ 128 x^4 r^2 (1-r^2)^2
-   8 x^3 r^2 (1-r^2) 
\nonumber\\&&\times
    (22 + 5 r^2+ 2 r^4)
-   4 x^2 r^2 (1-r^2) (46 - 13 r^2 + 8 r^4)
- 2 x (156 - 56 r^2 
\nonumber\\&&
       + 4 r^4 + 38 r^6 -13 r^8 )
- (2-r^2) ( 156 - 28 r^2 - 24 r^4 + 19 r^6 )
               \big]^2 ,
\\
  j_{0+}(x)&=&
(r^2/8) (1+x)^2 \big[
   64 x^5 r^2 (1-r^2)^2
- 128 x^4 r^2 (1-r^2)^2
-   8 x^3 r^2 (1-r^2)
\nonumber\\&&\times
    ( 1 + 26 r^2+ 2 r^4)
+   4 x^2 r^2 (1-r^2) (52 - 31 r^2+ 8 r^4)
- 2 x (156 + 28 r^2
\nonumber\\&&
        - 164 r^4 + 143 r^6 - 34 r^8 )
+(2-r^2) ( 156 - 40 r^2 + 18 r^4 + r^6 )
               \big]^2,
\\
  j_{++}(x)&=&
(r^4/4) (1-x^2) \big[
  32 x^5 r^2 (1-r^2)^2
-  4 x^3 r^2 (1-r^2) (  57 - 30 r^2 +2 r^4 )
- 24 x^2 r^2 
\nonumber\\&&\times
(1-r^2)
- x (156 - 196 r^2 + 284 r^4 - 137 r^6  + 22 r^8)
+ 6 r^2 (2-r^2)^2  
                \big]^2.
\end{eqnarray}
\label{ji-def}
\end{subequations}
The total ratio $R$ is obtained by summing over the helicities 
$\lambda_1$ and $\lambda_2$ and integrating $x$ from $-1$ to $1$.

We have not calculated the cross section for $J/\psi + \psi_2(1D)$. 
It receives contributions only from the nonfragmentation diagrams in 
Figs.~\ref{fig1}(c) and \ref{fig1}(d).
We therefore expect it to be comparable to 
the cross section for $\eta_c + \eta_{c2}(1D)$.

\section{ Predictions for $\bm{B}$ factories
	\label{sec:Bfactories}}

In this section, we calculate the cross sections for exclusive
double-charmonium production in $e^+ e^-$ annihilation at the $B$
factories. We also analyze the errors in the prediction for the $J/\psi
+ J/\psi$ production cross section.

\subsection{ Cross sections
	\label{sec:sigma}}

The results in Section~\ref{sec:CSM} were expressed in terms of the
ratio $R$, which is defined in Eq.~(\ref{R-def}). The corresponding
cross sections are
\begin{eqnarray}
\sigma[ H_1+H_2] = \frac{4 \pi \alpha^2}{3 s} R[ H_1+H_2].
\end{eqnarray}
The ratios $R$ depend on 
the coupling constant $\alpha$, the charm-quark mass $m_c$,
and the NRQCD matrix elements $\langle O_1 \rangle_H$.

\begin{table}[t]
\caption{\label{tab:O}%
The NRQCD matrix elements $\langle O_1 \rangle_H$. $H$ is a charmonium
state. These matrix elements are computed in Ref.~\cite{Braaten:2002fi}
from electromagnetic annihilation decays, using $m_c = 1.4$ GeV. The
errors are the statistical errors associated with the experimental inputs
only.}
\begin{ruledtabular}
\begin{tabular}{l|cccc}
$H$
& $\eta_c$, $J/\psi$
& $\eta_c(2S)$, $\psi(2S)$
& $h_c$(1P), $\chi_{cJ}(1P)$
& $\eta_{c2}(1D)$, $\psi_1(1D)$
\\
\hline
$\langle O_1\rangle_H$
& 0.335 $\pm$ 0.024 GeV$^3$
& 0.139 $\pm$ 0.010 GeV$^3$
& 0.053 $\pm$ 0.009 GeV$^5$
& 0.095 $\pm$ 0.015 GeV$^7$
\end{tabular}
\end{ruledtabular}
\end{table}

For the charm-quark mass $m_c$, we use the next-to-leading-order pole
mass, which can be expressed in terms of the running mass $\bar m_c(\bar
m_c)$ as
\begin{eqnarray}
m_c = \bar m_c(\bar m_c) \left(1 + \frac{4}{3} \frac{\alpha_s}{\pi} \right).
\end{eqnarray}
If we take the running mass of the charm quark to be
$\bar m_c(\bar m_c) = 1.2 \pm 0.2$~GeV, 
then the pole mass is $m_c = 1.4 \pm 0.2$~GeV.

The NRQCD matrix element $\langle O_1 \rangle_H$ can be determined
phenomenologically from the electromagnetic annihilation decay rate of
either $H$ or of a state that is related to $H$ by the heavy-quark spin
symmetry. We make use of the values of the matrix elements that were
obtained from such an analysis in Ref.~\cite{Braaten:2002fi}. That
analysis takes into account the next-to-leading-order QCD corrections to
the annihilation rate in the cases of the $S$- and $P$-wave states. It
obtains the NRQCD matrix elements for the $2S$ and $1D$ states from the
electronic decay rates of the $\psi(2S)$ and $\psi_1(1D)$, under the
assumption that the mixing between the $2S$ and $1D$ states is negligible.
The values of the matrix elements from that analysis are summarized in
Table~\ref{tab:O} for the case $m_c = 1.4$ GeV. For different values of
$m_c$, the values in Table~\ref{tab:O} should be multiplied by $(m_c/1.4
\, {\rm GeV})^{2+2L}$, where $L=0$, 1, and 2 for $S$-waves, $P$-waves,
and $D$-waves, respectively.

Our predictions for double-charmonium cross sections are given in
Table~\ref{tab:sigma--} for $C=-1$ states and in Table~\ref{tab:sigma++}
for $C=+1$ states. The error bars are those associated with the
uncertainty in the pole mass $m_c$ only.  The small error bars for the
$\eta_c + \eta_c$ cross section in Table~\ref{tab:sigma++} are a
consequence of the value of $m_c$ being fortuitously close to a zero in
the derivative of the cross section with respect to $m_c$. Note that the
cross sections in Table~\ref{tab:sigma--} are in fb, while those in
Table~\ref{tab:sigma++} are in units of $10^{-3}$~fb. All of the cross
sections, except those for which both of the charmonia are in $1^{--}$
states, are at least 3 orders of magnitude smaller than the $J/\psi +
J/\psi$ cross section. The cross sections for the $1^{--}$ states are
dominated by the photon-fragmentation diagrams in Figs.~\ref{fig1}(a)
and \ref{fig1}(b). For $m_c=1.4$~GeV, the photon-fragmentation 
diagrams contribute 114.5\% of the
$J/\psi+J/\psi$ cross section. The nonfragmentation diagrams in
Figs.~\ref{fig1}(c) and \ref{fig1}(d) contribute 0.9\%, while the
interference term contributes $-$15.4\%.

\begin{table}[t]
\caption{\label{tab:sigma--}%
Cross sections in fb for $e^+ e^-$ annihilation at $E_{\rm beam}=5.3$~GeV
into double-charmonium states $H_1+H_2$ with $C =-1$. The errors are
only those from variations in the pole mass $m_c = 1.4 \pm 0.2$~GeV.
There are additional large errors associated with perturbative and
relativistic corrections, as discussed in the text.}
\begin{ruledtabular}
\begin{tabular}{l|cccc}
$H_2$ $\backslash$ $H_1$
&   $J/\psi$          &     $\psi(2S)$      &   $h_c(1P)$
                                           &   $\psi_1(1D)$      \\
\hline
$J/\psi$
& 6.65 $\pm$ 3.02 & 5.52 $\pm$ 2.50 & (6.1 $\pm$ 0.9)$\times10^{-3}$
                                          & 0.80 $\pm$ 0.32\\
$\psi(2S)$
&                    & 1.15 $\pm$ 0.52 &(2.5 $\pm$ 0.4)$\times10^{-3}$
                                          & 0.33 $\pm$ 0.13 \\
$h_c(1P)$
&                    &                    &(0.19 $\pm$ 0.02)$\times10^{-3}$
                                          & \\
\end{tabular}
\end{ruledtabular}
\end{table}
\begin{table}
\caption{\label{tab:sigma++}%
Cross sections in units of $10^{-3}$~fb for $e^+ e^-$ annihilation at 
$E_{\rm beam}=5.3$~GeV into double-charmonium states $H_1+H_2$ with $C
=+1$. The errors are only those from variations in the pole mass $m_c
= 1.4 \pm 0.2$~GeV.
There are additional large errors associated with perturbative and
relativistic corrections, as discussed in the text.}
\begin{ruledtabular}
\begin{tabular}{l|cccccc}
$H_2$ $\backslash$ $H_1$
&$\eta_c$ 
&$\eta_c(2S)$ 
&$\chi_{c0}(1P)$
&$\chi_{c1}(1P)$
&$\chi_{c2}(1P)$
&$\eta_{c2}(1D)$       \\
\hline
$\eta_c$
& 1.83 $\pm$ 0.10     
& 1.52 $\pm$ 0.08     
& 0.34 $\pm$ 0.04     
& 1.31 $\pm$ 0.29     
& 0.48 $\pm$ 0.10     
& 0.18 $\pm$ 0.02   \\
$\eta_c(2S)$
&                                   
& 0.31 $\pm$ 0.02     
& 0.14 $\pm$ 0.02     
& 0.54 $\pm$ 0.12     
& 0.20 $\pm$ 0.04     
& 0.07 $\pm$ 0.01     
\end{tabular}
\end{ruledtabular}
\end{table}

\begin{figure}
\includegraphics[height=12cm]{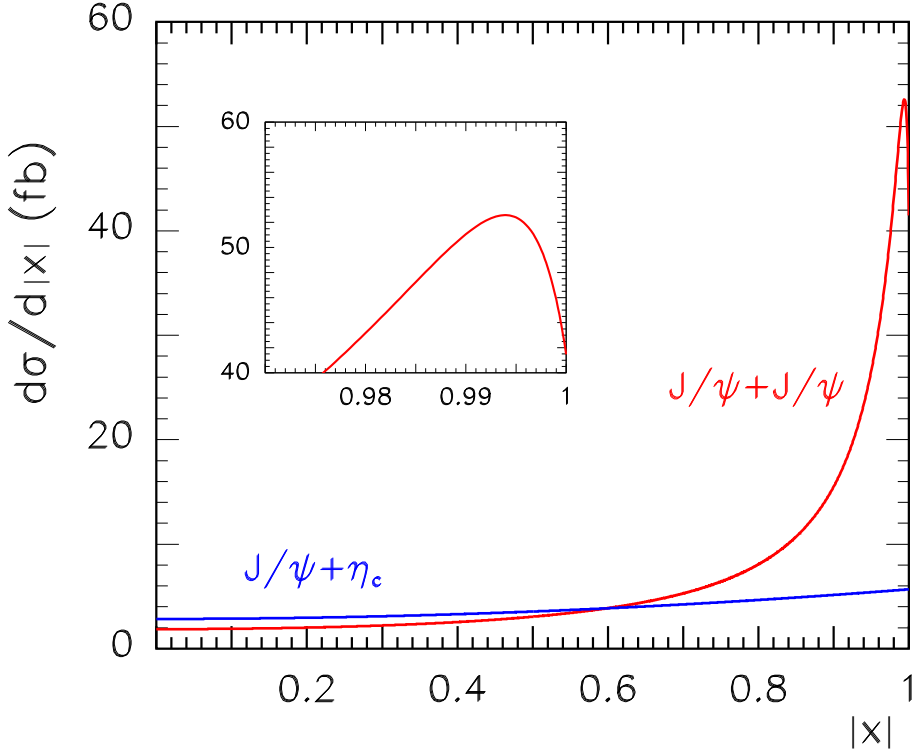}
\caption{\label{fig2}
Differential cross sections $d\sigma/d|x|$ for $e^+ e^-$ annihilation at
$E_{\rm beam}=5.3$~GeV into $J/\psi+J/\psi$ and $J/\psi+\eta_c$. 
The areas under the curves
are the integrated cross sections 6.65 fb and 3.78 fb.
There are large errors associated with perturbative and
relativistic corrections, as discussed in the text. }
\end{figure}

The cross sections for production of double-charmonium states with
opposite C-parity were calculated in
Refs.~\cite{Braaten:2002fi,Liu:2002wq}. The cross sections at the
$B$-factory energy in Table II of Ref.~\cite{Braaten:2002fi} were
calculated using the same NRQCD matrix elements as the cross section in
Table~\ref{tab:O}. In spite of the suppression factor of
$\alpha^2/\alpha_s^2$, the predicted cross section for $J/\psi+J/\psi$
is larger than that for $J/\psi+\eta_c$ by about a factor of 1.8. The
kinematic enhancement factor that scales as $r^{-6}$ more than
compensates for the suppression factor of $\alpha^2/\alpha_s^2$. Four
of the powers of $r^{-1}$ come from the enhancement associated with
fragmentation processes.  The other two powers of $r^{-1}$ come from the
fact that the final state $J/\psi(\pm 1)+\eta_c$ violates hadron
helicity conservation by one unit. The angular distributions $d
\sigma/d|x|$ for $m_c = 1.4$ GeV are shown in Fig.~\ref{fig2}. The
normalizations are such that the areas under the curves are equal to the
integrated cross sections 6.65 fb and 3.78 fb. At $x=0$, the
differential cross section for $J/\psi+J/\psi$ (normalized as in
Fig.~2) is smaller than that for
$J/\psi+\eta_c$ by about a factor 0.66. However the differential cross
section for $J/\psi+J/\psi$ is strongly peaked near the beam direction
at $x = \pm 0.994$, where it is larger than that for $J/\psi+\eta_c$ by
about a factor 9.3. The reason for the sharp peak is evident from the
expression (\ref{dRdx:frag}) for the limiting behavior as $r\to 0$ with
$x$ fixed. As we have already mentioned, this peak in the angular 
distribution leads to an enhancement factor $\ln(8/r^4)$ in the 
integrated cross section.

\subsection{ Perturbative corrections
	\label{sec:radiative}}

The perturbative corrections to the cross section for $J/\psi + J/\psi$
have not yet been calculated. However the perturbative corrections to
the dominant photon-fragmentation diagrams in Figs.~\ref{fig1}(a) and
\ref{fig1}(b) are closely related to the perturbative correction to the
electromagnetic annihilation decay rate for $J/\psi \to e^+ e^-$, which
gives a multiplicative factor
\begin{eqnarray}
\left( 1 - \frac{8}{3} \, \frac{\alpha_s}{ \pi} \right)^2.
\label{gam-psi}
\end{eqnarray}
The perturbative correction to the photon-fragmentation terms in the
cross section for $J/\psi + J/\psi$ is just the square of the expression
(\ref{gam-psi}). If we choose the QCD coupling constant to be $\alpha_s
= 0.25$, which corresponds to a renormalization scale $2 m_c$, then the
perturbative correction yields a multiplicative factor $(0.79)^4=0.39$.
The same perturbative correction factor applies to the cross sections
for $J/\psi + \psi(2S)$ and $\psi(2S) + \psi(2S)$. This perturbative
correction factor applies only to the leading contributions to the cross
sections in the limit $r \to 0$. However, since these contributions are
dominant, we conclude that the perturbative corrections are likely to
decrease the cross sections by about a factor of 3.

\subsection{ Relativistic corrections
	\label{sec:Rel}}

The prescription for calculating the leading relativistic correction for
$S$-wave charmonium production processes was summarized in
Ref.~\cite{Braaten:2002fi}. The leading relativistic correction is
conveniently expressed in terms of a quantity $\langle v^2 \rangle_H$
that can be defined formally as a ratio of matrix elements in NRQCD. It
can also be determined phenomenologically, using the mass $M_H$ of the
charmonium state as input \cite{Gremm:1997dq}.

If we keep only the dominant photon-fragmentation contribution to the
cross section for $J/\psi + J/\psi$, then the relativistic
correction to the cross section is closely related to the relativistic
correction to the electromagnetic annihilation decay rate for $J/\psi
\to e^+ e^-$. The relativistic correction to the photon-fragmentation
probability $P_{\gamma \to J/\psi}$ in Eq.~(\ref{P-gampsi}) is a
multiplicative factor
\begin{eqnarray}
\left( 1 - \frac{1}{ 6} \, \langle v^2 \rangle_H \right)^2
\times \left( \frac{2 m_c}{M_{J/\psi}} \right)^3.
\label{rel-corr}
\end{eqnarray}
The first factor in the expression (\ref{rel-corr}) arises from the
expansion of the electromagnetic current in powers of $v$
\cite{Braaten:1998au}, while the second factor arises from the
dependence of the rate on the charmonium mass, which can be determined
from dimensional analysis. The decay rate for $J/\psi \to e^+ e^-$ is
multiplied by the same factor, except that $2 m_c/M_{J/\psi}$ is raised
to the power 2 instead of 3. The relativistic correction associated with
the phase space for $J/\psi + J/\psi$ can be taken into account by
replacing the factor $(1-r^2)^{1/2}$ in Eq.~(\ref{dRdx:psi+psi}) by the
expression for $P_{\rm CM}/E_{\rm beam}$ given by Eq.~(\ref{P_cm}). If
all of these correction factors are taken into account, both in the
decay rate for $J/\psi \to e^+ e^-$ and in $P_{\gamma \to J/\psi}$, then
the net effect on the prediction for the production cross section for
$J/\psi + J/\psi$ is simply to multiply it by
\begin{eqnarray}
\left(\frac{2 m_c}{M_{J/\psi} }\right)^2
\times \frac{P_{\rm CM}/E_{\rm beam}}{(1-r^2)^{1/2}}.
\label{rel-fac}
\end{eqnarray}
For $m_c = 1.4$ GeV, the relativistic corrections decrease the cross
sections for $J/\psi+J/\psi$, $J/\psi+\psi(2S)$, and $\psi(2S)+\psi(2S)$
by multiplicative factors of about 0.78, 0.62, and 0.49, respectively.
The relativistic corrections also decrease the sensitivity to the value
of $m_c$.  The error associated with variations of $m_c$ in the cross
section for $J/\psi+J/\psi$ in Table~\ref{tab:sigma--} is about 41\%
for $m_c = 1.4$ GeV. After we take into account the relativistic
correction in (\ref{rel-fac}), this error decreases to about 9\%.

The relativistic corrections to the $J/\psi+J/\psi$ cross section are
significantly smaller than and have the opposite sign from the
relativistic corrections to the $J/\psi+\eta_c$ cross section, which are
given in Ref.~\cite{Braaten:2002fi}. For $m_c = 1.4$ GeV, the
relativistic corrections to the $J/\psi+\eta_c$ cross section are
estimated to increase the cross section by about a factor 5.5. The large
difference in the relativistic corrections suggests that there may be
large relativistic corrections not only to the absolute cross sections
for double-charmonium production, but also to the ratios of those cross
sections.

\subsection{ Phenomenology
	\label{sec:predict}}

Recently, the Belle Collaboration has measured the cross section for the
production of $J/\psi + \eta_c$ by observing a peak in the momentum
spectrum of the $J/\psi$ that corresponds to the 2-body final state
$J/\psi + \eta_c$ \cite{Abe:2002rb}. The measured cross section is
\begin{eqnarray}
\sigma[J/\psi+\eta_c] \times B[\ge 4] 
= \left( 33^{+7}_{-6} \pm 9 \right) \; {\rm fb},
\label{Belle}
\end{eqnarray}
where $B[\ge 4]$ is the branching fraction for the $\eta_c$ to decay
into at least 4 charged particles.  Since $B[\ge 4]<1$, the right side
of Eq.~(\ref{Belle}) is a lower bound on the production cross section
for $J/\psi + \eta_c$. This lower bound is about an order of magnitude
larger than the predictions of NRQCD in the nonrelativistic limit
\cite{Braaten:2002fi,Liu:2002wq}. The relativistic corrections are
large, and they may account for part of the discrepancy
\cite{Braaten:2002fi}. There may also be large nonperturbative
corrections to double-charmonium production at the $B$-factory energy
\cite{Liu:2002wq}. However the large discrepancy between the predictions
and the measurement is still disturbing. In the Belle fit to the $J/\psi$
momentum distribution, the full width at half maximum of the $\eta_c$
peak is about 0.11 GeV. Since the mass difference between the $J/\psi$
and $\eta_c$ is about 0.12 GeV, there are probably $J/\psi + J/\psi$
events that contribute to the $J/\psi + \eta_c$ signal that is observed
by Belle. If these were taken into account, they would increase the
compatibility between the NRQCD prediction of the cross section and the
Belle measurement.

The signal for the production of $J/\psi +X$, with $m_X$ near the mass
of the $\eta_c$, can be resolved into the contributions from
$J/\psi+\eta_c$ and from $J/\psi+J/\psi$.  One way to do this is to
measure the angular distribution of the production products and fit to a
linear combination of the angular distributions in Fig.~\ref{fig2}. This
method tests not only the predicted production rates, but also the
dominance of the photon-fragmentation production mechanism. 
Alternatively, one can measure directly the rate for the production of
$J/\psi+J/\psi$ with subsequent decay of both $J/\psi$'s into lepton
pairs. In the analysis of Ref.~\cite{Abe:2002rb}, the data set for
$J/\psi + X$, with $m_X$ near the mass of the $\eta_c$, contains only
$67 \pm 13$ events.  Since the branching fraction for $J/\psi$ into
$e^+e^-$ or $\mu^+\mu^-$ is only about 0.12, that data sample may not be
large enough to allow one to employ this method.  However, as the integrated
luminosity increases, the double-dilepton signal should become
observable.

The Belle Collaboration also saw evidence for $J/\psi + \chi_{c0}(1P)$
and $J/\psi + \eta_c(2S)$ events. The $\eta_c(2S)$ was recently
discovered by the Belle Collaboration at a mass of $M_{\eta_c(2S)} =
3654 \pm 6 \pm 8$ MeV~\cite{Choi:2002na}. Since the mass difference
between the $\psi(2S)$ and the $\eta_c(2S)$ is only about 0.03 GeV, any
signal for $J/\psi + \eta_c(2S)$ in the $J/\psi$ momentum spectrum is
probably also contaminated by $J/\psi + \psi(2S)$ events. A 3-peak fit
to the momentum spectrum of the $J/\psi$ by the Belle Collaboration
gives approximately 67, 39, and 42 events with an accompanying
$\eta_c$, $\chi_{c0}(1P)$, or $\eta_c(2S)$, with an uncertainty of 12-15
events. The predictions in Ref.~\cite{Braaten:2002fi} for the
relative cross sections for the production of a $J/\psi$ with an
accompanying $\eta_c$, $\chi_{c0}(1P)$, or $\eta_c(2S)$ are 1.00,
0.63, and 0.42, respectively. The observed proportion of events is
compatible with the NRQCD predictions. 

In summary, we have calculated the cross sections for $e^+ e^-$
annihilation through two virtual photons into exclusive
double-charmonium states. The cross sections are particularly large if
the two charmonia are both $1^{--}$ states. In the absence of radiative
and relativistic corrections, the predicted cross section for the
production of $J/\psi + J/\psi$ at the $B$ factories is larger than that
for $J/\psi + \eta_c$ by a factor of about 3.7. The perturbative and
relativistic corrections for these two processes may be rather different
and could significantly change the prediction for the ratio of the cross
sections. Nevertheless, the inclusion of contributions from processes
involving two virtual photons in the theoretical prediction for the
cross section for $J/\psi + \eta_c$ production is likely to decrease the
large discrepancy between that prediction and the Belle measurement.

\begin{acknowledgments}
One of us (E.B.) would like to thank B.~Yabsley for valuable discussions.
The research of G.T.B.~and J.L.~in the High Energy Physics Division at
Argonne National Laboratory is supported by the U.~S.~Department of
Energy, Division of High Energy Physics, under Contract W-31-109-ENG-38.
The research of E.B.~is supported in part by the U.~S.~Department of
Energy, Division of High Energy Physics, under grant DE-FG02-91-ER4069
and by Fermilab, which is operated by Universities Research Association
Inc.~under Contract DE-AC02-76CH03000 with the U.~S.~Department of
Energy.
\end{acknowledgments}



\begin{thebibliography}{}

\bibitem{Abe:2002rb}
K.~Abe {\it et al.}  [Belle Collaboration],
Phys.\ Rev.\ Lett.\  {\bf 89}, 142001 (2002)
[arXiv:hep-ex/0205104].

\bibitem{BBL}
G.~T.~Bodwin, E.~Braaten, and G.~P.~Lepage,
	Phys.\ Rev.\ D {\bf 51}, 1125 (1995);
	{\bf 55}, 5853(E) (1997)
	[arXiv:hep-ph/9407339].

\bibitem{CSM}
M.~B.~Einhorn and S.~D.~Ellis,
Phys.\ Rev.\ D {\bf 12}, 2007 (1975);
S.~D.~Ellis, M.~B.~Einhorn, and C.~Quigg,
Phys.\ Rev.\ Lett.\  {\bf 36}, 1263 (1976);
C.~E.~Carlson and R.~Suaya,
Phys.\ Rev.\ D {\bf 14}, 3115 (1976);
J.~H.~K\"uhn,
Phys.\ Lett.\ B {\bf 89}, 385 (1980);
T.~A.~DeGrand and D.~Toussaint,
Phys.\ Lett.\ B {\bf 89}, 256 (1980);
J.~H.~K\"uhn, S.~Nussinov, and R.~R\"uckl,
Z.\ Phys.\ C {\bf 5}, 117 (1980);
M.~B.~Wise,
Phys.\ Lett.\ B {\bf 89}, 229 (1980);
C.-H.~Chang,
Nucl.\ Phys.\ B {\bf 172}, 425 (1980);
R.~Baier and R.~R\"uckl,
Phys.\ Lett.\ B {\bf 102}, 364 (1981);
E.~L.~Berger and D.~Jones,
Phys.\ Rev.\ D {\bf 23}, 1521 (1981);
W.~Y.~Keung,
in {\it The Cornell $Z^0$ Theory Workshop},
edited by M.~E. Peskin and S.-H. Tye
(Cornell University, Ithaca, 1981). 

\bibitem{Braaten:2002fi}
E.~Braaten and J.~Lee,
Phys.\ Rev.\ D~(to be published),
arXiv:hep-ph/0211085.

\bibitem{Liu:2002wq}
K.-Y.~Liu, Z.-G.~He, and K.-T.~Chao,
arXiv:hep-ph/0211181.

\bibitem{Bodwin:2002fk}
G.~T.~Bodwin, J.~Lee, and E.~Braaten,
Phys.\ Rev.\ Lett.~(to be published),
arXiv:hep-ph/0212181.

\bibitem{Chernyak:dj}
V.~L.~Chernyak and A.~R.~Zhitnitsky,
Sov.\ J.\ Nucl.\ Phys.\  {\bf 31}, 544 (1980)
[Yad.\ Fiz.\  {\bf 31} (1980) 1053].

\bibitem{Brodsky:1981kj}
S.~J.~Brodsky and G.~P.~Lepage,
Phys.\ Rev.\ D {\bf 24}, 2848 (1981).

\bibitem{Fleming:1994iu}
S.~Fleming,
Phys.\ Rev.\ D {\bf 50}, 5808 (1994)
[arXiv:hep-ph/9403396].

\bibitem{Chang:1997dw}
C.-H.~Chang, C.-F.~Qiao, and J.-X.~Wang,
Phys.\ Rev.\ D {\bf 56}, 1363 (1997)
[arXiv:hep-ph/9704364];
Phys.\ Rev.\ D {\bf 57}, 4035 (1998).

\bibitem{Gremm:1997dq}
M.~Gremm and A.~Kapustin,
Phys.\ Lett.\ B {\bf 407}, 323 (1997)
[arXiv:hep-ph/9701353].


\bibitem{Braaten:1998au}
E.~Braaten and Y.-Q.~Chen,
Phys.\ Rev.\ D {\bf 57}, 4236 (1998);
{\bf 59}, 079901(E) (1999)
[arXiv:hep-ph/9710357].

\bibitem{Choi:2002na}
S.-K.~Choi {\it et al.}  [Belle Collaboration],
	Phys.\ Rev.\ Lett.\  {\bf 89}, 102001 (2002);
	{\bf 89}, 129901(E) (2002)
	[arXiv:hep-ex/0206002].


\end{thebibliography}
\end{document}